# Theoretical investigation on electronic properties and carrier mobilities of armchair graphyne nanoribbons

Hongyu Ge, Guo Wang[*] and Yi Liao

*Department of Chemistry, Capital Normal University, Beijing 100048, China*

[*]E-mail: wangguo@mail.cnu.edu.cn

Seven types of armchair graphyne nanoribbons are investigated with HSE06 functional. The quantum confinements in the graphyne nanoribbons open or increase the band gaps of the corresponding two-dimensional graphynes, which is crucial to high on/off ratio in electronic device operation. The major carrier mobilities of the graphyne nanoribbons with high percentage of *sp* hybridized carbon atoms are very large. The sparse linking pattern results in small number of frontier crystal orbitals and small deformation potential constants, which are responsible for the large carrier mobilities. Some graphyne nanoribbons have band gaps larger than 0.4 eV. Meanwhile, they have both high hole and electron mobilities. These benefit current complementary circuit with low power dissipation. Especially, the hole and electron mobilities of 14,14,18-graphyne nanoribbons are more than an order larger than those of the armchair graphene nanoribbons, indicating that they have potential applications in high speed electronic devices.

## Introduction

For a long time, carbon was thought to have only two crystalline forms, diamond and graphite. With the discovery of $C_{60}$, carbon nanotube and graphene, the form of carbon covers all dimensions (from zero to three-dimension). The era of carbon allotropes [1] brings us many exciting properties, such as super-hardness and high performance in field effect transistors. Beyond these well-known allotropes, many other carbon forms especially for two-dimensional sheet are predicted [2-5], such as graphynes (GYs) [6] and graphdyines (GDYs) [7] The rich geometrical variety makes carbon a promising material for various potential applications.

Two-dimensional GY [6], which can be constructed by replacing some carbon-carbon bonds in graphene with acetylenic linkages, recently received great interests for its potential competition for graphene [2]. The mixing of *sp* and *sp*$^2$ hybridized carbon atoms produces a serial of GY structures. First principal calculations indicated that some GYs have versatile Dirac cones. Their electronic properties are even more amazing than that of graphene [8]. Furthermore, theoretical investigation based on deformation potential theory revealed that the carrier mobilities of 6,6,12-graphyne should be even higher than those of graphene [9].

Carrier mobility is one of the most important parameter in electronic devices, and is essential for high speed operation. Graphene [10] is one of the most potential candidates for future ultra-fast electronic devices [11, 12]. The carrier mobilities can be up to $10^5$ cm$^2$V$^{-1}$s$^{-1}$ at room temperature [13]. Theoretical investigations [14] also indicated that graphene has both high hole and electron mobilities with the order of $10^5$ cm$^2$V$^{-1}$s$^{-1}$. Despite of the exciting properties, electronic devices made of graphene can be hardly switched off due to zero band gaps. Modern complementary circuit with low power dissipation requires not only high hole and electron mobilities, but also a finite band gap, preferably larger than 0.4 eV [15]. It is encouraging that finite width of one-dimensional graphene nanoribbons can introduce band gaps. However the carrier mobilities are significantly degraded [16]. Like graphene, GYs with Dirac cones have zero band gaps, which should also be hardly switched off. With the scale of electronic industry down and down, one-dimensional nanoribbons should be important because they can produce narrower devices and the current in circuit is essentially one-dimensional. It is important to make clear that whether GY nanoribbons have considerably large band gaps, and whether GY nanoribbons have both high hole and electron carrier mobilities.

In this work, several one-dimensional armchair GY nanoribbons are constructed and investigated by using density functional theory. It is indicated that some GY nanoribbons have considerable large band gaps (larger than 0.4 eV). Some GY nanoribbons have both high hole and electron mobilities. The reason for the high carrier mobilities is closely related to their linking pattern, and is explained with crystal orbital analysis.

**Models and computational details**

GY nanoribbons shown in Figure 1 are constructed from the two-dimensional sheets GY1-GY7 [6], which are also called γ-graphyne and 6,6,12-graphyne, 6,6,14-graphyne, β-graphyne, 14,14,14-graphyne, 14,14,18-graphyne and α-graphyne, respectively. One-dimensional GY nanoribbons derived from these GYs should have armchair or zigzag edges. However, it is indicated that some zigzag GY1, GY2, GY4 and GY7 nanoribbons are magnetic [17, 18], which should be useful in other fields. Their frontier bands do not have dispersions as good as those of armchair graphene nanoribbons [19]. The latter ones have light carrier masses which is essential for high carrier mobilities. For this reason, armchair GY nanoribbons are focused in the present work. The armchair nanoribbons are named $N$-GY ($N$=1-8), in which $N$ represents the width of the nanoribbons denoted in Figure 1. Unlike graphene, GY can not be exfoliated from any naturally occurring material. A synthetic method should be a solution. In fact, a similar system GDY has been synthesized via a cross-coupling reaction by using organic reagents [20]. Therefore, the edges of GY nanoribbons are passivated by hydrogen atoms, which should be already presented in reactants.

A band gap is an important parameter for electronic devices, because a sufficiently large band gap is essential to achieve large on/off ratio [15]. The screened hybrid density functional HSE06 [21] is used, which can accurately describe the band gaps for solids [22]. Full geometrical optimization of the structures are performed with Bloch functions based on a standard 6-21G($d$, $p$) basis set in CRYSTAL14 program [23, 24] Atomic grid with 55 radial and 434 angular points is used in density functional numerical integration. Default values of convergence criteria are used. Monkhorst-Pack samplings with more than 40 k-points per Å$^{-1}$ in the first Brillouin zone are used. The k-point net is sufficiently dense to obtain the converged geometries and related properties of the structures. When calculating the band structures, the k-point sampling are 20 times denser in order to facilitate the fitting of carrier effective mass at a certain range. It is noted that without a sharp density of states near frontier band edges, carriers in a range wider than $k_B T$ should participate in the conduction in electronic devices. It this work, the energy range 10 $k_B T$ [25] is used to obtain the effective masses.

In the deformation potential theory [26], the wavelength of an electron is estimated to be much larger than the lattice constant. For the GY nanoribbons with band gaps larger than 0.4 eV, the velocities near frontier band edges are in the range of 0.4-1.9×10$^5$ ms$^{-1}$. The corresponding de Broglie wavelengths are in the range of 38-172 Å, which are also much larger than the lattice constants of the GY nanoribbons. According to the deformation potential theory for semiconductors [26], the carriers in the GY nanoribbons are mostly scattered by longitudinal acoustic phonons with long wavelengths. The carrier mobilities of the one-dimensional GY nanoribbons with considerable band gaps are obtained by [27]

$$\mu_{1D} = \frac{e\hbar^2 C_{1D}}{(2\pi k_B T)^{1/2} |m^*|^{3/2} E_1^2},$$

in which $C_{1D}$ is a stretching modulus for one-dimensional crystals, $m^*$ is a carrier effective mass and $E_1$ is a deformation potential constant. Carrier effective masses, stretching moduli, and deformation potential constants are needed in order to calculate carrier mobilities. Carrier effective masses are obtained via fitting frontier bands, while stretching moduli and deformation potential constants are calculated under deformed geometries [28]. The deformation potential theory is successfully applied to various systems, such as carbon nanotube [29, 30], graphene [14, 31] and MoS$_2$ [32].

## Results and discussions

### Structures and stabilities

All the GY nanoribbons have planar structures. The optimized lattice constants for $N$-GY1 to $N$-GY7 ($N$ = 1-8) are 6.87, 6.88, 6.88, 9.45-9.46, 6.93-7.01, 9.48-9.54, and 12.03-12.09 Å, respectively. For GY1-GY3 nanoribbons, the difference of the lattice

constants is less than 0.01 Å. For GY4-GY7 nanoribbons, the lattice constant gradually decreases with the increasing width. Nevertheless, the lattice constants of all the GY nanoribbons with large width should approach the values 6.87, 6.88, 6.88, 9.45, 6.90, 9.47 and 12.03 Å for the two-dimensional GYs along armchair directions, which are calculated with the same method for comparison.

GY1 is mostly investigated two-dimensional GY, and its lattice constant is in a narrow range of 6.86-6.90 Å in different calculations [2]. Our result based on HSE06 functional is 6.87 Å, which is in agreement with LDA [33], GGA [34-36] and even with MNDO calculated results [6]. This implies that the geometry is not very sensitive to the calculation methods in this case. Single-layer graphene is very light with a density of 0.76 mg per m$^2$, according to the optimized geometry with the same method. For two-dimensional GY1-GY7, the densities are 77%, 73%, 73%, 61%, 66%, 56% and 50% as large as that of graphene, respectively. GYs are even lighter than graphene because they have bigger carbon rings and thus bigger holes than graphene has.

Since the ratio of carbon and hydrogen atoms is different for different GY nanoribbons, it is difficult to compare the stabilities by directly comparing their energies. A Gibbs free energy of formation per atom defined as [37]

$$\delta G(x) = -E_c(x) - x\mu_C - (1-x)\mu_H$$

is used to investigate the relative stabilities for these binary systems $C_xH_{1-x}$. In the equation, $E_c$ is the cohesive energy per atom of the GY nanoribbons, $\mu_C$ and $\mu_H$ are chemical potentials of carbon and hydrogen, respectively. We choose $\mu_C$ and $\mu_H$ as $-E_c$(graphene) and $-E_c$(H$_2$) per atom. For two-dimensional GYs, the free energies can be also calculated from the equation, where $x$ equals to 1.

The Gibbs free energies shown in Figure 2 are all positive (from 0.46-1.05 eV), indicating that they are less stable than graphene. From two-dimensional GY1 to GY7, the stability generally decreases with increasing percentage of acetylenic linkages. The exceptions are GY3 and GY5, whose free energies are about 0.01 and 0.03 eV lower than those of GY2 and GY4, respectively. However, the differences are quite small. When acetylenic linkages are introduced into GYs, the *sp* hybridized carbon atoms connect to only two atoms rather than three for the $sp^2$ hybridized ones in graphene. The connection in GYs is not as sufficient as that in graphene, and the percentage of $\pi$ bonds is also larger in GYs. Since a $\pi$ bond is not as strong as a $\sigma$ bond, the stability decreases for GYs. The percentages of the $sp$ hybridized carbon atoms are 50%, 56%, 58%, 67%, 67%, 71% and 75% for GY1-GY7 [6] The percentage increases except that GY4 and GY5 have the same percentage (67%). Furthermore, the percentage difference between GY2 (56%) and GY3 (58%) is small. The different linking patterns should be related to the small energetic exceptions for GY3 and GY5. For GY1, GY5 and GY7, the stability also decreases from DF-TB calculations [38].

The relative stabilities of the GY nanoribbons with the same $N$ are similar to those of the two-dimensional GYs. The free energies of the GY nanoribbons are lower than those of the corresponding GYs as shown in Figure 2, indicating that the GY nanoribbons are more thermodynamically stable. The stability decreases with the width for all types of the GY nanoribbons. The narrowest GY nanoribbons have the smallest free energies, which are 0.11-0.26 eV lower than those of the two-dimensional GY1-GY7. It is noticed that the percentages of the $sp$ hybridized carbon atoms are slightly smaller than those of the GYs. The narrowest nanoribbons are taken as examples. For 1-GY1 to 1-GY7, these values are 40%, 46%, 50%, 67%, 62%, 69% and 72%, respectively. The percentages are less than those of the GYs, except for only 1-GY4, whose value (67%) is equal to that of GY4. The energy difference between 1-GY4 and GY4 is 0.11 eV, which is the smallest among all the energy difference between the 1-GYs and two-dimensional GYs. The lower percentage of $sp$ hybridized carbon atoms should contribute to the stability of the GY nanoribbons. Although the stabilities of the GYs and GY nanoribbons are lower than graphene, it does not imply that they can not be synthesized. In fact, GDY with adjacent acetylenic linkages was synthesized [20]. Stable carbon atomic chains are also experimentally realized [39], which has the highest percentage (100%) of $sp$ hybridized carbon atoms.

**Electronic properties**

The calculation based on high level HSE06 functional indicates that two-dimensional GY1 has a direct band gap at M point shown in Figure 3(a). The valence band maximum (VBM) and conduction band minimum (CBM) are labeled with cycles. The band gap is 1.07 eV, which is larger than 0.52 eV [33, 40] or 0.47 eV [34] calculated with pure density functionals and is comparable to the value 1.2 eV based on extended Huckel theory with a correction. Unlike the geometries described above, band gaps are sensitive to the methods adopted. Since a band gap is an important parameter which affects on/off ratio in electronic device operation and power consumption, accurately describing the band gaps is necessary. For GY3, direct band gap (0.37 eV) exists at X' point shown in Figure 3(b). For GY2, GY4 and GY7, Dirac points exist in the band structures, which are consistent with previous results [8, 9]. Only the band structures of GY7 are shown in Figure 3(c) for brevity. There is a gap between the lowest unoccupied (LU) and LU+1 band. This might be useful when turning off the GY7-based electronic device by gate voltage. However, the gap is only 0.23 eV. Besides GY5 [38], GY6 also has metallic band structures shown in Figure 3(d), in which two frontier bands cross at the Fermi level. It is shown that different linking patterns result in different electronic properties. GY1 and GY3 are semiconductors, GY2, GY4 and GY7 are zero-gap semiconductors with Dirac cones, while GY5 and GY6 are metals. According to the standard 0.4 eV for semiconductor devices [15], only GY1 is good for low power dissipation electronic devices.

All the GY nanoribbons shown in Figure 1 have direct band gaps. For GY2, GY3, GY5 and GY7 nanoribbons, VBM and CBM are at $\Gamma$ point of the first Brillouin zone, while the two points are at X point for GY6 nanoribbons. Only the band structures of the narrowest GY nanoribbons are shown in Figure 4(c), 4(d), 4(h) and 4(j). For 1-GY1, the VBM and CBM are near X point as shown in Figure 4(a). For other GY1 nanoribbons, the VBM and CBM are at X point as shown in Figure 4(b). The situation for GY4 nanoribbons is a little complex. For 1-GY4, 5-GY4 and 7-GY4, the VBM and CBM are near X point as shown in Figure 4(e), while 4-GY4 has a direct band gap at X point in Figure 4(g). For other GY4 nanoribbons, the VBM and CBM are at $\Gamma$ point in Figure 4(f). Nevertheless, the band gap difference at different points for

GY4 nanoribbons is not large, and the band gaps are all less than 0.2 eV except for 1-GY4.

Compared with the two-dimensional GYs, the band gaps increase or become non-zero because of quantum confinements. The band gaps of GY1 nanoribbons shown in Figure 5(a) are in the range of 1.19-1.95 eV, which are all larger than that of the two-dimensional semiconducting GY1. The band gap decreases with the width and should reach 1.07 eV for GY1 with infinite width. For GY3 nanoribbons, the band gap decreases generally with the width. The band gaps of 7-GY3 and 8-GY3 are 0.39 eV, which is close to the value 0.37 eV for semiconducting GY3. For the other five zero-gap GYs, the band gap opens when the GY nanoribbons are formed. For GY2 nanoribbons, the band gap decreases from 0.96 to 0.16 eV for 1-GY2 to 3-GY2, while the band gaps are less than 0.1 eV for 4-GY2 to 7-GY2. For other GY nanoribbons, the band gap generally decreases with the width. Oscillation occurs especially for GY5, GY6 and GY7 nanoribbons shown in Figure 5(b). The variation is more complex than that for graphene nanoribbons with a uniform linking pattern. All the GY1 nanoribbons have band gaps larger than 0.4 eV, because the band gap of GY1 itself is already quite large. For GY3 with a band gap of 0.37 eV, $N$-GY3 ($N$=1-6) meet the standard. For other GYs with zero band gaps, the quantum confinement in GY nanoribbons makes 1-GY2, 1-GY4, 1-GY5, 3-GY5, 1-GY6, 2-GY6, 4-GY6 and 2-GY7 have band gaps larger than 0.4 eV. The GY nanoribbons provide more possibilities for electronic devices, because only GY1 has a band gap larger than 0.4 eV among the two-dimensional GYs.

**Carrier mobilities**

Carrier mobilities are calculated under deformation potential theory [26] for structures with band gaps larger than 0.4 eV. The one-dimensional stretching moduli $C_{1D}$ increase with the width of the GY nanoribbons, for there are more atoms in the wider GY nanoribbon. The structures are stronger for wider GY nanoribbons. The stretching moduli increase almost linearly with the width. The slopes are 16.4, 14.2, 14.0, 12.7, 6.5, 8.3 and 9.4 eVÅ$^{-2}$, respectively. These values are close to the in-plane stiffness of the two-dimensional GYs along the armchair directions. Generally, the mechanical strength decreases with the percentage of the *sp* hybridized carbon atoms, or with the density of GY1-GY7. GY5 and GY6 with metallic properties are two exceptions, which have weaker strength than other semiconductors. Since the carrier mobilities depend linearly on the $C_{1D}$ and the difference between different GY nanoribbons is not large, the influence on the carrier mobilities coming from the stretching moduli is not very crucial.

A deformation potential constant equals to an energy change at VBM or CBM with respect to lattice deformation. For GY1 nanoribbons, all the valence band deformation potential constant $E_{1v}$ (4.90-6.58 eV) are larger than the conduction band deformation potential constant $E_{1c}$ (1.03-1.52 eV) listed in Table 1. The hole and electron effective masses for each GY1 nanoribbons are similar (0.23-0.35 $m_0$), so the electron mobilities of GY1 nanoribbons are all larger than the hole mobilities. Unlike the stretching moduli, deformation potential constants significantly affect the carrier mobilities. The electron mobilities are more than an order larger than the hole mobilities. With the increasing stretching modulus with the width, the mobilities increase gradually. The largest electron mobility is 31643 $cm^2V^{-1}s^{-1}$ for the widest 8-GY1. It is indicated that the electron mobility of the two-dimensional GY1 ($10^4$ $cm^2V^{-1}s^{-1}$) is also larger than the hole mobility [41]. The GY1 and GY1 nanoribbons are suitable for electron transport. Similar situation occurs for GDY and GDY nanoribbons [42, 43].

For *N*-GY3 (*N*=1-6), the deformation potential constants especially for $E_{1c}$ decreases compared with GY1 nanoribbons. The $E_{1c}$ are as small as 0.82-0.98 eV. Carrier effective masses (0.16-0.19 $m_0$) are also lower than those of GY1 nanoribbons. The two aspects make GY3 nanoribbons have higher electron mobilities than GY1 nanoribbons have. The carrier mobilities generally increase with the width. For *N*-GY3 (*N*=4-6), the electron mobilities reach the order of $10^5$ $cm^2V^{-1}s^{-1}$. The hole mobilities are in the order of $10^3$ $cm^2V^{-1}s^{-1}$ and are all smaller than the electron mobilities due to the larger $E_{1v}$ (5.12-5.55 eV). Similar to GY1 nanoribbons, GY3 nanoribbons are favorable to electron transport.

GY2, GY4 and GY7 have zero band gaps with Dirac cones. Only one nanoribbon for each two-dimensional structure has band gap larger than 0.4 eV. This is different from the situation for GY1 and GY3, which are already finite band gap semiconductors. 1-GY2 has similar deformation potential constants and carrier masses with 1-GY1, so the carrier mobilities are also comparable to 1-GY1. However, the polarity is inversed. The hole mobility (4460 cm$^2$V$^{-1}$s$^{-1}$) is an order larger than the electron mobility (276 cm$^2$V$^{-1}$s$^{-1}$). For 1-GY4, although the electron mass (0.47 $m_0$) is slightly bigger than that of 1-GY1, the $E_{1c}$ is much smaller (0.56 eV). The electron mobility (16092 cm$^2$V$^{-1}$s$^{-1}$) is about twice as large as that of 1-GY1 (8394 cm$^2$V$^{-1}$s$^{-1}$). For 2-GY7, because of the small $E_{1v}$ (0.19 eV) and small hole mass (0.17 $m_0$), the hole mobility is up to 7×10$^5$ cm$^2$V$^{-1}$s$^{-1}$. Unlike GY1 and GY3 nanoribbons, 1-GY2 and 2-GY7 are favorable to hole transport. However there is only one nanoribbon with a considerable large band gap for each GY, they do not possess both high hole and electron mobilities.

GY5 and GY6 are metals. Confinement in one-dimension completely changes their electronic properties. 1-GY5 and 3-GY5 have band gaps larger than 0.4 eV. 1-GY5 has similar $E_{1c}$ (0.67 eV) with 1-GY4, but much smaller electron effective mass (0.15 $m_0$), so its electron mobility (36254 cm$^2$V$^{-1}$s$^{-1}$) is higher than 1-GY4. The polarity of 3-GY5 is different from that of 1-GY5. The hole mobility of 3-GY5 reaches 10$^4$ cm$^2$V$^{-1}$s$^{-1}$. GY5 nanoribbons with considerable large band gaps have both high hole and electron mobilities, which are essential for modern complementary circuit. GY6 nanoribbons are even better than GY5 nanoribbons. There are three nanoribbons 1-GY6, 2-GY6 and 4-GY6 with band gaps larger than 0.4 eV. For 1-GY6 and 4-GY6, the major carrier (carrier with larger mobility) is hole. Because of their small $E_{1v}$ (0.30 and 0.38 eV) and small effective masses (0.14 and 0.16 $m_0$), the hole mobilities are 281366 and 296090 cm$^2$V$^{-1}$s$^{-1}$, reaching the order of 10$^5$ cm$^2$V$^{-1}$s$^{-1}$. These values are much larger than the electron mobilities (932 and 1623 cm$^2$V$^{-1}$s$^{-1}$). For 2-GY6, because of the small $E_{1c}$ (0.10 eV), the electron mobility is as high as 4588971 cm$^2$V$^{-1}$s$^{-1}$, which is in the order of 10$^6$ cm$^2$V$^{-1}$s$^{-1}$. Two GY5 nanoribbons and three GY6 nanoribbons with considerable large band gaps have both high hole and electron mobilities, which should be useful for high speed complementary circuit.

From GY1 nanoribbons to GY7 nanoribbons, the percentage of *sp* hybridized carbon atoms increases generally. The atoms are much sparse in GY7 nanoribbons than in GY1 nanoribbons. The average carrier effective masses are 0.30, 0.32, 0.18, 0.37, 0.21, 0.14, 0.17 $m_0$ for the seven GY nanoribbon. Compared with GY1 nanoribbons, GY3, GY5, GY6 and GY7 nanoribbons have smaller effective masses. There is only one nanoribbon with a considerable large band gap for GY2 and GY4. Also the VBM and CBM are at a special point (near X point) for 1-GY4, so the average masses for GY1 and GY4 nanoribbons do not decrease. The average deformation potential constants for major carriers are 1.22, 1.36, 0.92, 0.56, 1.05, 0.26 and 0.19 for GY1 to GY7 nanoribbons. The deformation potential constants decrease except for GY2 and GY4 nanoribbon also because there is only one structure for each nanoribbon.

Armchair graphene nanoribbons with hydrogen-passivated edges (AGNR) are also calculated with the same method for comparison. The polarity alternates with the width, which is in agreement with previous results [19, 31]. The stretching moduli also increase linearly with the width. The slope is 25.3 eVÅ$^{-2}$, which is much larger than the largest value 16.4 eVÅ$^{-2}$ for GY1 nanoribbons. The major carrier mobilities increase with the width. The widest AGNR with a band gap (0.41 eV) larger than 0.4 eV is 34-AGNR. The width is 41 Å, which is comparable to the value 34 Å for 4-GY6. The $E_{1v}$ and $E_{1c}$ are 4.38 and 11.71 eV, which are much larger those of the GY nanoribbons. As shown in Figure 6(a) and 6(b), the highest occupied crystal orbital (HOCO) at VBM and the lowest unoccupied crystal orbital (LUCO) at CBM are parallel or vertical to the one-dimensional extended direction. The delocalized HOCO should has smaller energy change than the localized LUCO during the deformation along one-dimensional direction [19, 31], so $E_{1v}$ is smaller than $E_{1c}$ for 34-AGNR. For GY nanoribbons, the situation is similar. The narrowest GY nanoribbons are taken as examples. Only the frontier orbitals for major carriers are shown in Figure 6. The crystal orbitals are distributed on all the carbon atoms except the *sp* hybridized carbon atoms, which have parallel triple bonds along the one-dimensional direction. GY4, GY6 and GY7 nanoribbons with large eighteen-membered rings have small average deformation potential constants 0.56, 0.26 and 0.19 eV. Those of other GY nanoribbons are 1.22, 1.36, 0.92 and 1.05 eV, which are around or larger than 1 eV. The atoms in GY nanoribbons with large eighteen-membered rings are very sparse. The crystal orbitals are also sparse in these GY nanoribbons. This is directly related to the percentage of *sp* hybridized carbon atoms as well as the density of the GY nanoribbons. During the deformation, the energy change for sparse crystal orbital should be smaller than that for dense crystal orbitals. This is the reason why GY6 and GY7 have very small deformation potential. Compared with GY nanoribbons, AGNRs have rather dense linking pattern, which have much larger deformation potential constants.

The small deformation potential constant should be responsible to the high carrier mobilities. The major carrier mobilities for GY1 and GY4 nanoribbons reach $10^4$ $cm^2V^{-1}s^{-1}$. Some GY3 nanoribbons and GY7 nanoribbon have the mobilities of $10^5$ $cm^2V^{-1}s^{-1}$. However, these nanoribbons are only favorable to one type of carrier transport. GY5 nanoribbons have major carrier mobilities of $10^4$ $cm^2V^{-1}s^{-1}$, while GY6 nanoribbons reach $10^5$-$10^6$ $cm^2V^{-1}s^{-1}$. GY5 and GY6 nanoribbons have not only band gaps larger than 0.4 eV, but also both high hole and electron mobilities, which should be helpful for high speed complementary circuit.

For graphene nanoribbons, although 34-AGNR has larger stretching modulus and smaller carrier effective masses (0.10 $m_0$), the deformation potential constant for major carrier of 34-AGNR is much larger than those of the GY nanoribbons. The major carrier mobility is 15327 $cm^2V^{-1}s^{-1}$. This value is comparable to those of GY1 nanoribbons, but is much smaller than those GY6 and GY7 nanoribbons. Like AGNR, GY6 nanoribbons have both high hole and electron mobilities depending on their width, but the major mobilities are more than one order larger than those of the AGNRs.

## Conclusions

Seven types of GY nanoribbons as well as the two-dimensional GYs with different linking patterns are investigated by using density functional theory. The connection of *sp* hybridized carbon atoms is not as sufficient as that of *sp*$^2$ hybridized carbon atoms. With the increasing percentage of *sp* hybridized carbon atoms, the density, stability and mechanical strength generally decrease. Two-dimensional GYs have versatile electronic properties. GY1 and GY3 are semiconductors, GY2, GY4 and GY7 are zero-gap semiconductors with Dirac cones, while GY5 and GY6 are metals. Only GY1 has a band gap larger than 0.4 eV, which is essential for high on/off ratio in electronic device operation. All the GY nanoribbons have larger band gaps than the corresponding GYs because of quantum confinements. This provides more possibilities for electronic devices application. The deformation potential constants for major carriers generally decrease with the percentage of *sp* hybridized carbon atoms, which is explain by less number of frontier crystal orbitals on the sparse atoms. The small deformation potential constant results in large carrier mobilities. Interestingly, some GY5 and GY6 nanoribbons have both high hole and electron mobilities. Meanwhile, they have considerable large band gaps. These are beneficial to current complementary circuit with low power dissipation. AGNRs with band gaps larger than 0.4 eV also have both high hole and electron mobilities. The major mobilities of the GY6 nanoribbons are more than an order larger than those of the AGNRs, indicating potential applications in electronic devices.

**References**

[1] A. Hirsch, *Nature materials*, 2010, **9**, 868.

[2] A. L. Ivanovskii, *Prog. Solid State Chem.*, 2013, **41**, 1.

[3] M. A. Hudspeth, B. W. Whitman, V. Barone and J. E. Peralta, *ACS Nano*, 2010, **4**, 4565.

[4] Q. Song, B. Wang, K. Deng, X. Feng, M. Wagner, J. D. Gale, K. Müllen and L. Zhi, *J. Mater. Chem. C*, 2013, **1**, 38.

[5] H. Lu and S.-D. Li, *J. Mater. Chem. C*, 2013, **1**, 3677.

[6] R. H. Baughman, H. Eckhardt and M. Kertesz, *J. Chem. Phys.*, 1987, **87**, 6687.

[7] M. M. Haley, S. C. Brand and J. J. Pak, *Angew. Chem. Int. Ed. Engl.*, 1997, **36**, 836.

[8] D. Malko, C. Neiss, F. Viñes and A. Görling, *Phys. Rev. Lett.*, 2012, **108**, 086804.

[9] J. Chen, J. Xi, D. Wang and Z. Shuai, *J. Phys. Chem. Lett.*, 2013, **4**, 1443.

[10] K. S. Novoselov, D. Jiang, F. Schedin, T. J. Booth, V. V. Khotkevich, S. V. Morozov and A. K. Geim, *Proc. Natl. Acad. Sci.*, 2005, **102**, 10451.

[11] K. S. Novoselov, A. K. Geim, S. V. Morozov, D. Jiang, M. I. Katsnelson, I. V. Grigorieva, S. V. Dubonos and A. A. Firsov, *Nature*, 2005, **438**, 197.

[12] Y. Zhang, Y. W. Tan, H. L. Stormer and P. Kim, *Nature*, 2005, **438**, 201.

[13] S. V. Morozov, K. S. Novoselov, M. I. Katsnelson, F. Schedin, D. C. Elias, J. A. Jaszczak and A. K. Geim, *Phys. Rev. Lett.*, 2008, **100**, 016602.

[14] J. Xi, M. Long, L. Tang, D. Wang and Z. Shuai, *Nanoscale*, 2012, **4**, 4348.

[15] F. Schwierz, *Nature Nanotechnol.*, 2010, **5**, 487.

[16] X. Li, X. Wang, L. Zhang, S. Lee and H. Dai, *Science*, 2008, **319**, 1229.

[17] G. Yu, Z. Liu, W. Gao and Y. Zheng, *J. Phys.: Condens. Matter*, 2013, **25**, 285502.

[18] X. N. Niu, D. Z. Yang, M. S. Si and D. S. Xue, *J. Appl. Phys.*, 2014, **115**, 143706.

[19] M.-Q. Long, L. Tang, D. Wang, L. Wang and Z. Shuai, *J. Am. Chem. Soc.*, 2009, **131**, 17728.

[20] G. Li, Y. Li, H. Liu, Y. Guo, Y. Li and D. Zhu, *Chem. Commun.*, 2010, **46**, 3256.

[21] A. V. Krukau, O. A. Vydrov, A. F. Izmaylov and G. E. Scuseria, *J. Chem. Phys.*, 2006, **125**, 224106.

[22] T. M. Henderson, J. Paier and G. E. Scuseria, *Phys. Status Solidi B*, 2011, **248**, 767.

[23] R. Dovesi, R. Orlando, A. Erba, C. M. Zicovich-Wilson, B. Civalleri, S. Casassa, L. Maschio, M. Ferrabone, M. De La Pierre, Ph. D'Arco, Y. Noël, M. Causà, M. Rérat and B. Kirtman, *Int. J. Quantum Chem.*, 2014, **114**, 1287.

[24] R. Dovesi, V. R. Saunders, C. Roetti, R. Orlando, C. M. Zicovich-Wilson, F. Pascale, B. Civalleri, K. Doll, N. M. Harrison, I. J. Bush, Ph. D'Arco, M. Llunell, M. Causà, Y. Noël, *CRYSTAL14 User's Manual*, University of Torino, Torino, 2014.

[25] G. Grimvall, *The Electron-Phonon Interaction in Metals*, North-Holland Publishing Company, Amsterdam, 1981.

[26] J. Bardeen and W. Shockley, *Phys. Rev.*, 1950, **80**, 72.

[27] E. G. Wilson, *J. Phys. C: Solid State Phys.*, 1982, **15**, 3733.

[28] G. Wang and Y. Huang, *J. Phys. Chem. Solids*, 2007, **68**, 2003.

[29] G. Wang and Y. Huang, *J. Phys. Chem. Solids*, 2008, **69**, 2531.

[30] B. Xu, Y. D. Xia, J. Yin, X. G. Wan, K. Jiang, A. D. Li, D. Wu and Z. G. Liu, *Appl. Phys. Lett.*, 2010, **96**, 183108.

[31] G. Wang, *Chem. Phys. Lett.*, 2012, **533**, 74.

[32] Y. Cai, G. Zhang, and Y.-W. Zhang, *J. Am. Chem. Soc.*, 2014, **136**, 6269.

[33] N. Narita, S. Nagai, S. Suzuki and K. Nakao, *Phys. Rev. B*, 1998, **58**, 11009.

[34] J. Zhou, K. Lv, Q. Wang, X. S. Chen, Q. Sun and P. Jena, *J. Chem. Phys.*, 2011, **134**, 174701.

[35] H. Zhang, M. Zhao, X. He, Z. Wang, X. Zhang and X. Liu, *J. Phys. Chem. C*, 2011, **115**, 8845.

[36] J. Kang, J. Li, F. Wu, S.-S. Li and J.-B. Xia, *J. Phys. Chem. C*, 2011, **115**, 20466

[37] T. Dumitrica, M. Hua and B. I. Yakobson, *Phys. Rev. B*, 2004, **70**, 241303.

[38] A. N. Enyashin and A. L. Ivanovskii, *Phys. Status Solidi B*, 2011, **248**, 1879.

[39] C. Jin, H. Lan, L. Peng, K. Suenaga and S. Iijima, *Phys. Rev. Lett.*, 2009, **102,** 205501.

[40] V. Coluci, S. Braga, S. Legoas, D. Galvão and R. Baughman, *Phys. Rev. B*, 2003, **68**, 035430.

[41] J. Xi, D. Wang, Y. Yi and Z. Shuai, *J. Chem. Phys.*, 2014, **141**, 034704.

[42] M. Long, L. Tang, D. Wang, Y. Li and Z. Shuai, *ACS Nano*, 2011, **5**, 2593.

[43] H. Bai, Y. Zhu, W. Qiao and Y. Huang, *RSC Adv.*, 2011, **1**, 768.

Table 1. Valence band deformation potential constant $E_{1v}$, conduction band deformation potential constant $E_{1c}$ (eV), hole effective mass $|m_h^*|$, electron effective mass $|m_e^*|$ ($m_0$), hole mobility $\mu_h$ and electron mobility $\mu_e$ (cm$^2$V$^{-1}$s$^{-1}$) of the nanoribbons.

|  | $E_{1v}$ | $E_{1c}$ | $|m_h^*|$ | $|m_e^*|$ | $\mu_h$ | $\mu_e$ |
|---|---|---|---|---|---|---|
| 1-GY1 | 6.58 | 1.03 | 0.35 | 0.31 | 166 | 8394 |
| 2-GY1 | 6.20 | 1.52 | 0.29 | 0.23 | 423 | 9402 |
| 3-GY1 | 5.44 | 1.25 | 0.32 | 0.30 | 570 | 11734 |
| 4-GY1 | 5.23 | 1.25 | 0.30 | 0.26 | 855 | 17946 |
| 5-GY1 | 5.17 | 1.14 | 0.32 | 0.29 | 1042 | 24887 |
| 6-GY1 | 5.01 | 1.14 | 0.30 | 0.27 | 1315 | 29387 |
| 7-GY1 | 5.06 | 1.14 | 0.32 | 0.29 | 1329 | 31466 |
| 8-GY1 | 4.90 | 1.25 | 0.30 | 0.27 | 1774 | 31643 |
| 1-GY2 | 1.36 | 6.10 | 0.34 | 0.29 | 4460 | 276 |
| 1-GY3 | 5.55 | 0.98 | 0.18 | 0.16 | 754 | 29737 |
| 2-GY3 | 5.55 | 0.87 | 0.18 | 0.16 | 1157 | 58008 |
| 3-GY3 | 5.28 | 0.93 | 0.19 | 0.17 | 1752 | 69099 |
| 4-GY3 | 5.28 | 0.82 | 0.19 | 0.17 | 2040 | 103749 |
| 5-GY3 | 5.12 | 0.93 | 0.19 | 0.17 | 2715 | 97197 |
| 6-GY3 | 5.28 | 0.98 | 0.19 | 0.18 | 2878 | 97943 |
| 1-GY4 | 4.78 | 0.56 | 0.27 | 0.47 | 497 | 16092 |
| 1-GY5 | 1.59 | 0.67 | 0.31 | 0.15 | 2224 | 36254 |
| 3-GY5 | 1.42 | 5.66 | 0.19 | 0.18 | 10320 | 735 |
| 1-GY6 | 0.30 | 5.61 | 0.14 | 0.13 | 281366 | 932 |
| 2-GY6 | 5.17 | 0.10 | 0.12 | 0.11 | 1560 | 4588971 |
| 4-GY6 | 0.38 | 5.23 | 0.16 | 0.16 | 296090 | 1623 |
| 2-GY7 | 0.19 | 4.90 | 0.17 | 0.16 | 721873 | 1134 |
| 34-AGNR | 4.38 | 11.71 | 0.10 | 0.10 | 15327 | 2051 |

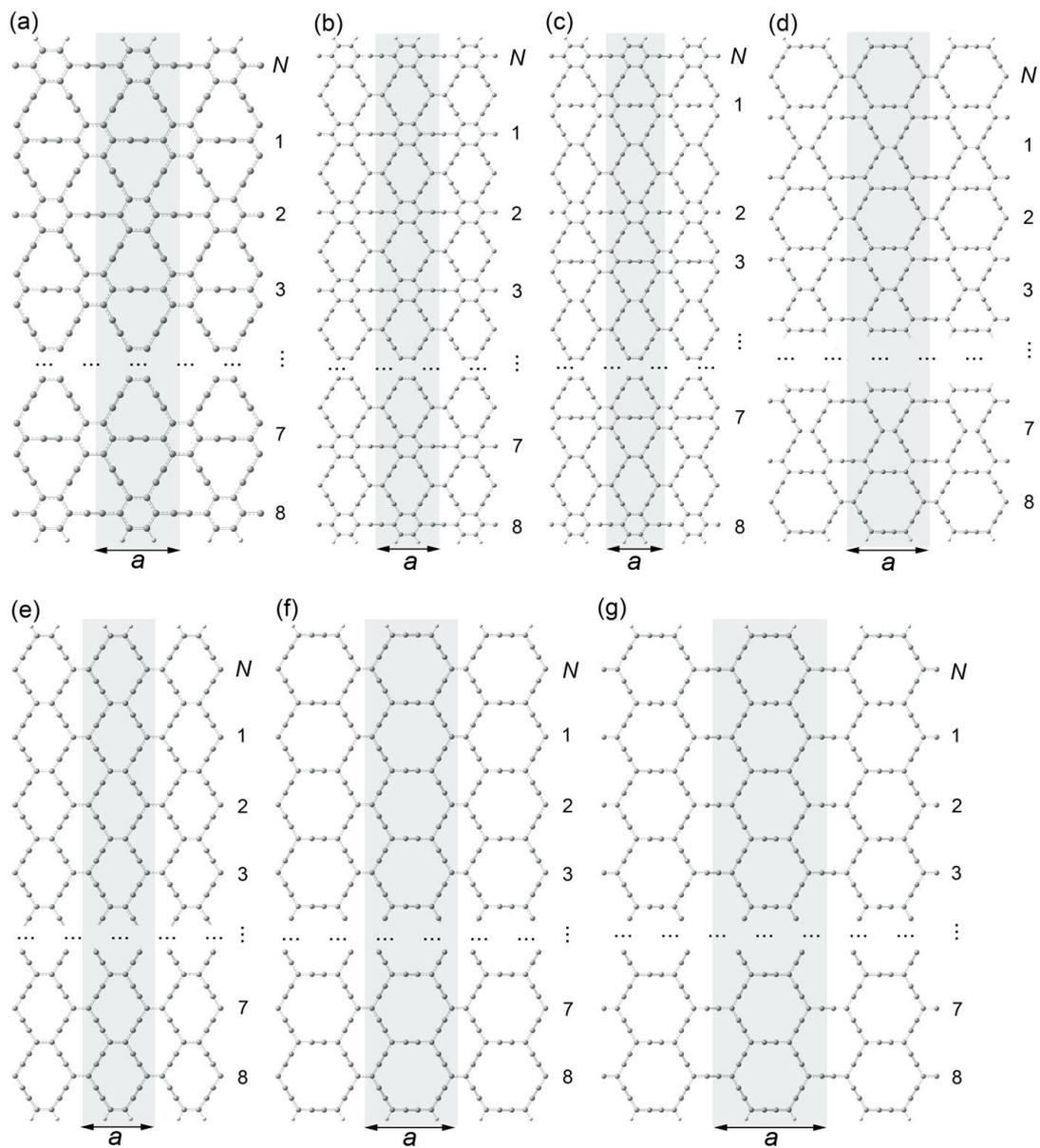

Figure 1. Models of GY nanoribbons. The unit cells are presented in shadow, and $a$ is the cell parameter.

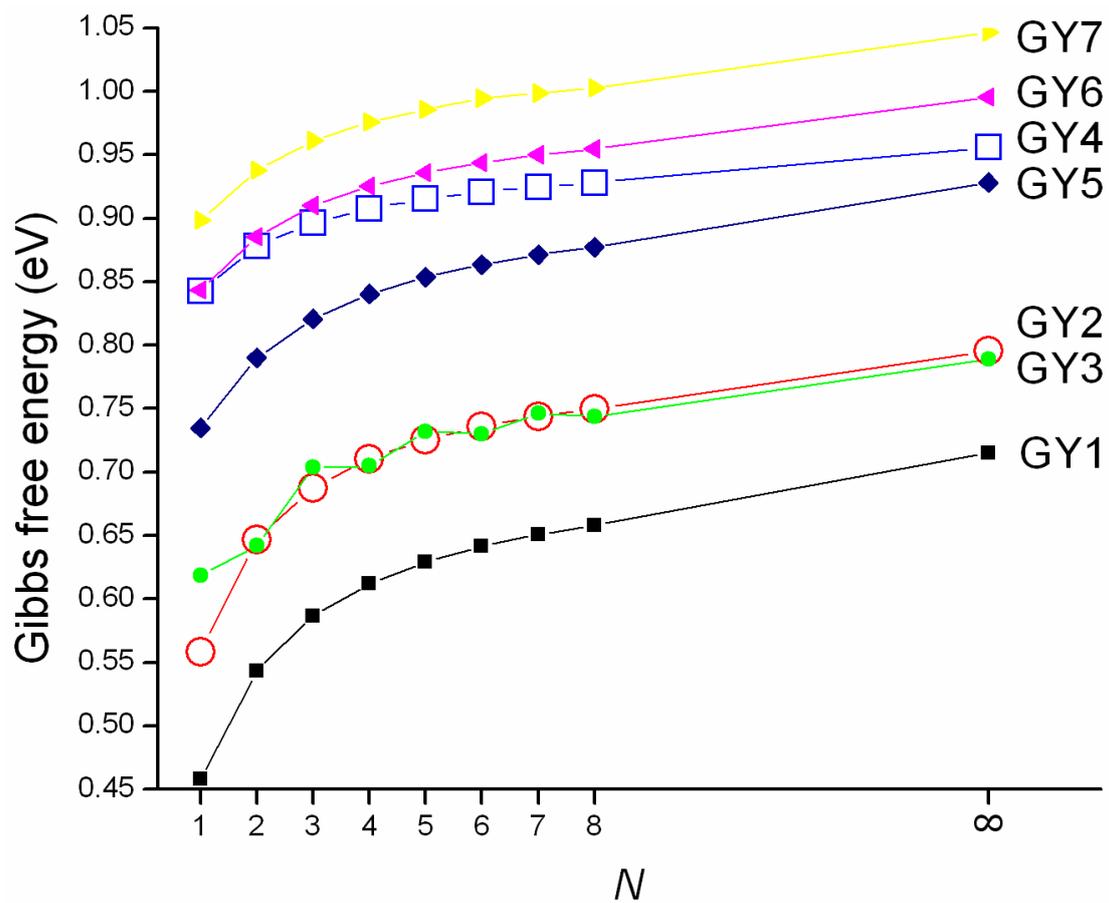

Figure 2. Gibbs free energies of *N*-GY nanoribbons (*N*=1-8) and GYs.

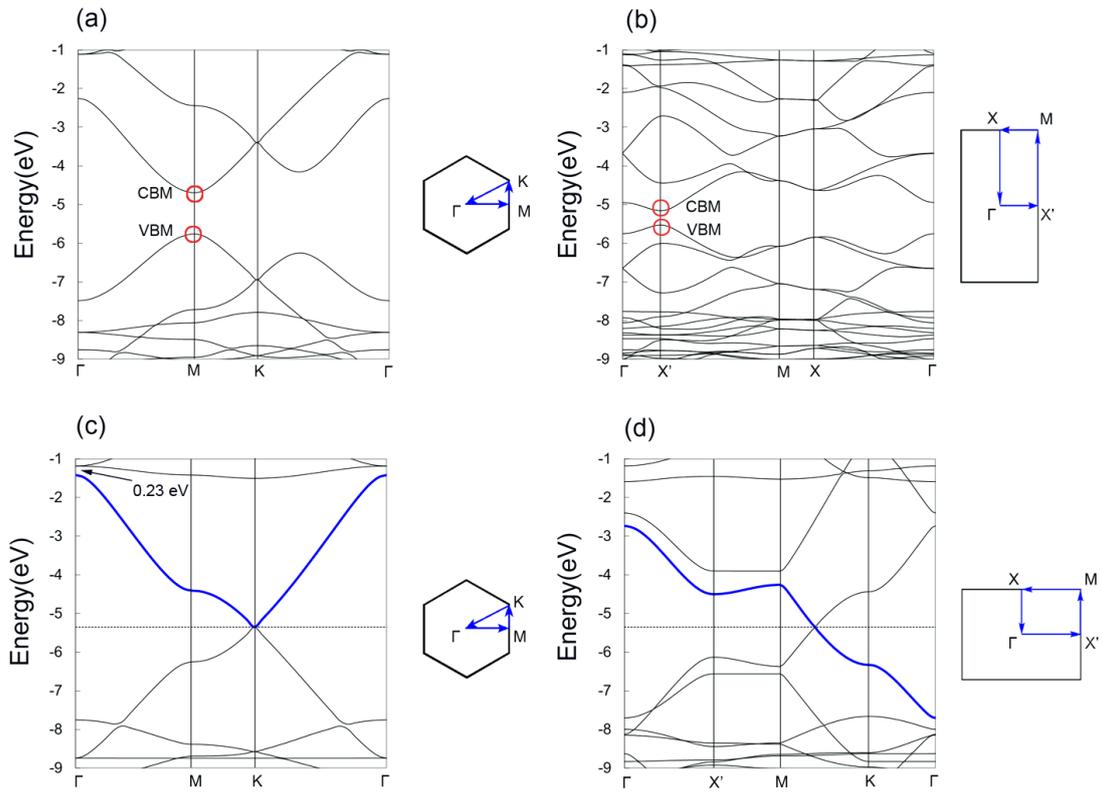

Figure 3. Band structures and first Brillouin zones of (a) GY1, (b) GY3, (c) GY7 and (d) GY6.

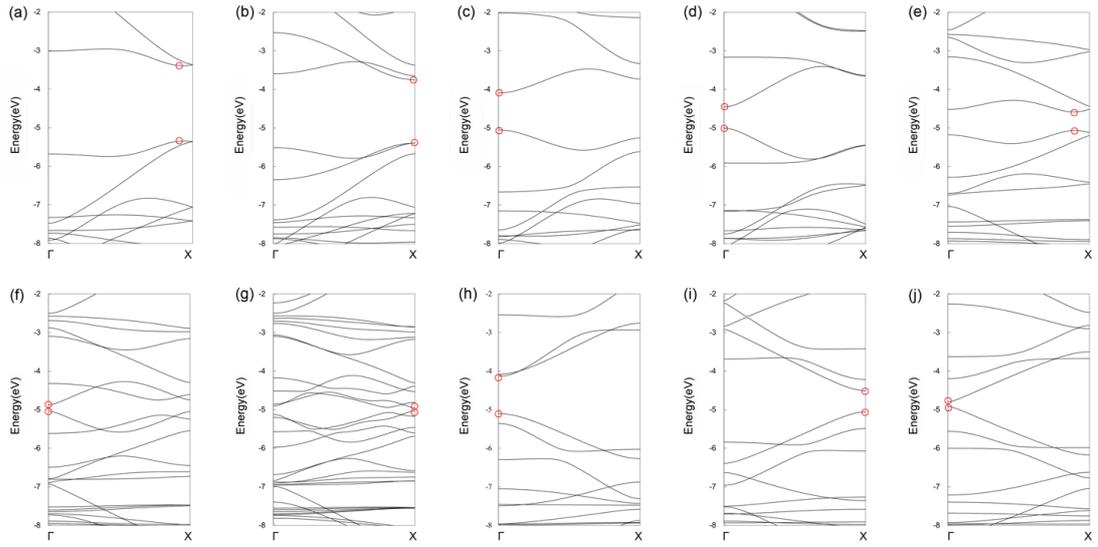

Figure 4. Band structures of (a) 1-GY1, (b) 2-GY1, (c) 1-GY2, (d) 1-GY3, (e) 1-GY4, (f) 2-GY4, (g) 4-GY4, (h) 1-GY5, (i) 1-GY6 and (j) 1-GY7.

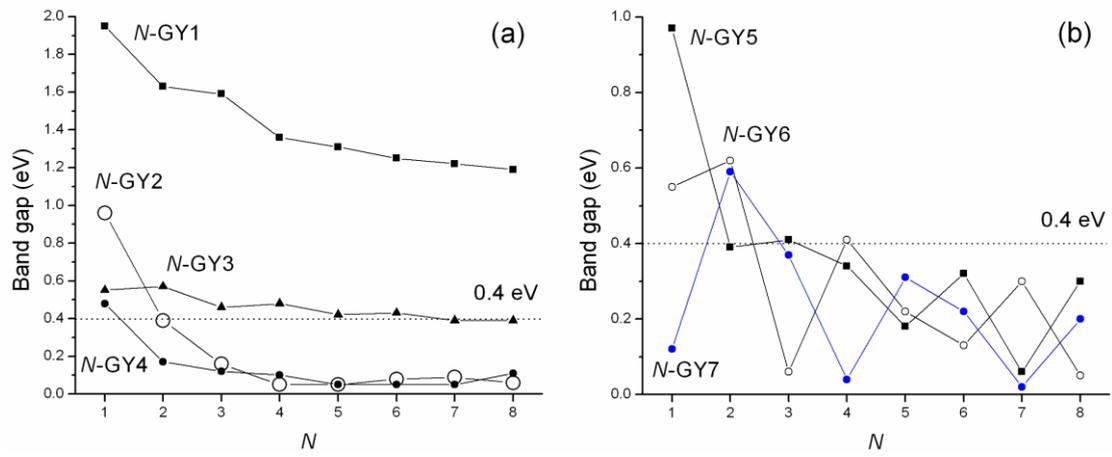

Figure 5. Band gaps of (a) GY1, GY2, GY3 and GY4 nanoribbons and (b) GY5, GY6 and GY7 nanoribbons.

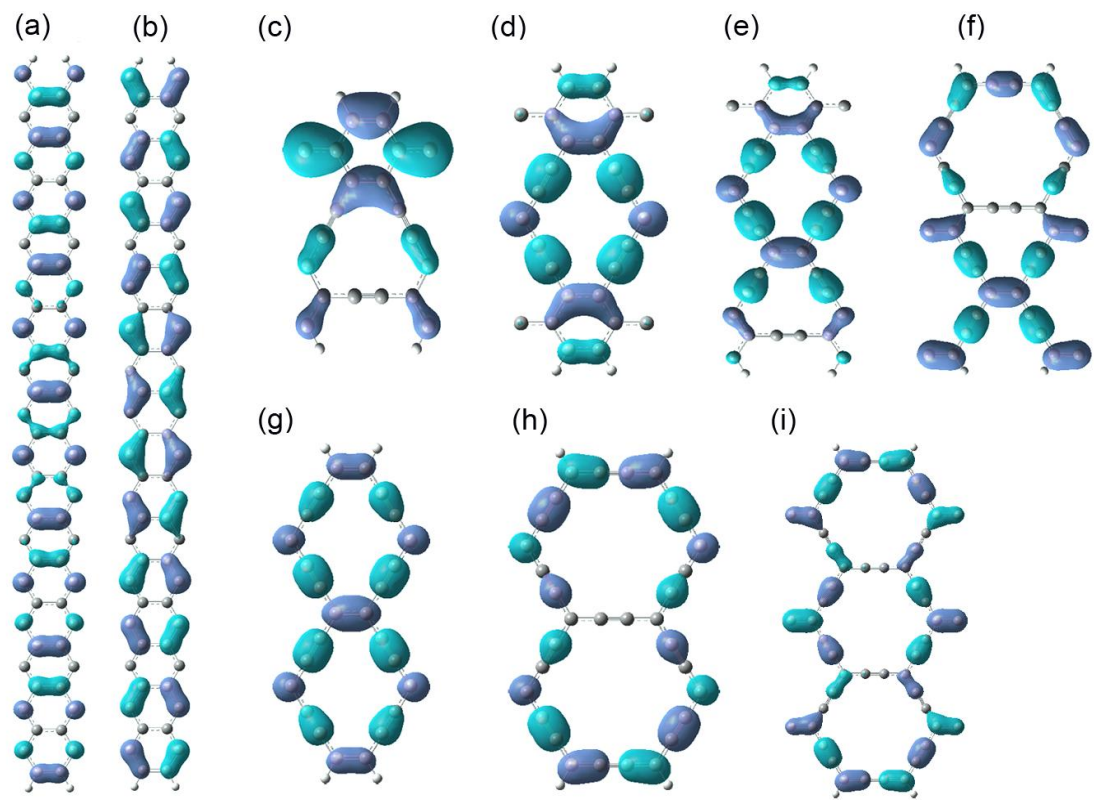

Figure 6. (a) HOCO of 34-AGNR, (b) LUCO of 34-AGNR, (c) LUCO of 1-GY1, (d) HOCO of 1-GY2, (e) LUCO of 1-GY3, (f) LUCO of 1-GY4, (g) LUCO of 1-GY5, (h) HOCO of 1-GY6, and (i) HOCO of 2-GY7.